# Search for a TeV Component of GRBs Using the Milagrito Detector

**Isabel R. Leonor**[1] **for the Milagro Collaboration**
[1]*Department of Physics and Astronomy, University of California at Irvine, Irvine, CA 92697, USA*

## Abstract

Observing gamma ray bursts (GRBs) in the TeV energy range can be extremely valuable in providing insight to GRB radiation mechanisms and in constraining source distances. The Milagrito detector was an air shower array which used the water Cherenkov technique to search for TeV sources. Data from this detector was analyzed to look for a TeV component of GRBs coincident with low energy $\gamma$-rays detected by the BATSE instrument on the Compton Gamma Ray Observatory. A sample of 54 BATSE GRBs which were in the field of view of the Milagrito detector during its lifetime (February 1997 to May 1998) was used.

## 1 Introduction

Gamma ray bursts are the most electromagnetically luminous objects observed in the universe, releasing energies of $10^{51}$ - $10^{53}$ ergs in a few seconds in the form of $\gamma$-rays (Piran, 1999). Over the past two years, considerable progress was made in detecting optical and x-ray counterparts to GRBs, which has led to confirmation of their cosmological origin and has provided valuable insight into GRB energy conversion and radiation mechanisms. Before the advent of such lower energy detections, GRB emission was detected exclusively in the 10 keV to 18 GeV energy range by instruments such as those on the Compton Gamma Ray Observatory (CGRO) satellite, one of which is the Burst and Transient Source Experiment (BATSE). BATSE has been extremely successful in the detection and study of gamma ray bursts in the soft $\gamma$-ray energy range of 10 keV to 100 MeV. Operating since 1991, it has been detecting gamma ray bursts at the rate of about one burst per day.

In spite of these observations and recent progress in detecting counterparts, the origin of GRBs − which were first observed about 30 years ago − remains an enigma. Opening a new energy window for GRB observation in the TeV regime will be an invaluable aid in solving this mystery. Detection of TeV emission from GRBs will: (1) provide valuable information for modelling GRB radiation mechanisms; (2) constrain general quantitative properties of GRBs such as Lorentz factor and shock radius (Pilla & Loeb, 1998); (3) provide an upper limit to source distances owing to the predicted absorption of TeV gamma rays by intergalactic infrared (IR) photons; (4) contribute to the accurate determination of GRB locations and source identification. Current observations and GRB models do not rule out the existence of such higher energy component of GRBs. Indeed, it has long been asked if the lack of observed hard $\gamma$-rays from GRBs is due to observational bias and the lack of detectors suitable for such observations.

Non-detection of a TeV component, though leading to more ambiguous conclusions, will also have an impact since this will − if one assumes the existence of a TeV component − set a lower limit on the source distance scale and will provide evidence for gamma-ray absorption either at the source or by IR photons. Since gamma-ray fluxes at these high energies are very low, ground-based detectors such as Milagrito are the relevant instruments for detecting GRBs at the TeV regime.

The Milagrito detector (McCullough, et al., 1999) was an air shower array which used the water Cherenkov technique to detect TeV gamma-ray sources. It was located in the Jemez mountains of New Mexico, near Los Alamos, at an altitude of 2650 m above sea level. It consisted of 228 photomultiplier tubes (PMTs) in a light-tight pond of water. Arrival times at these PMTs of water Cherenkov light from air shower particles were used to reconstruct cosmic ray or $\gamma$-ray events incident on the atmosphere. It was operational from February 1997 to May 1998, detecting extensive air showers at a rate of 300-400 $s^{-1}$. Its angular resolution was typically $\sim 1.0^o$. With its large field of view, high duty cycle, and a sensitivity to $\gamma$-rays of $\sim 1$ TeV, Milagrito was a practical instrument for observing gamma ray bursts which lasted but fleetingly and which were unpredictable in both time and position of occurrence.

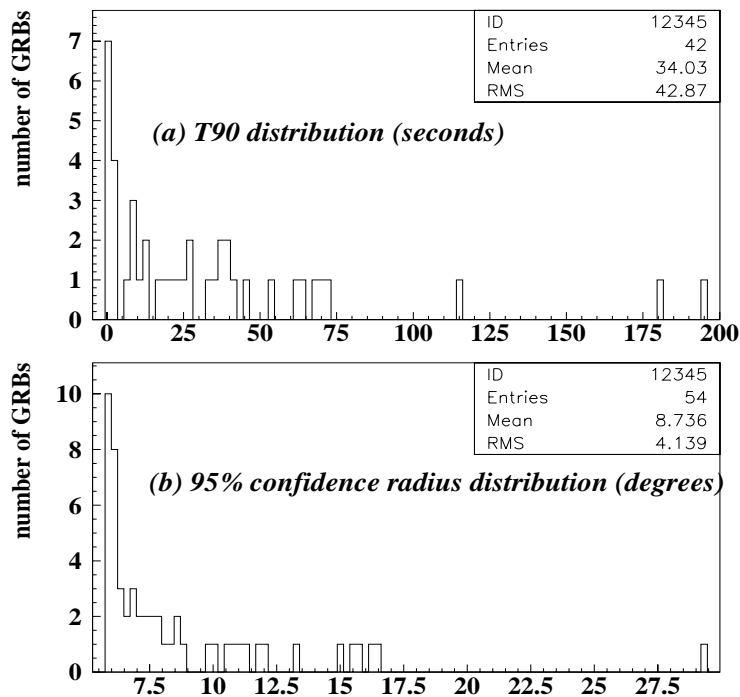

Figure 1: (a) T90 times and, (b) 95% confidence radii for the GRB sample.

During the Milagrito lifetime, 54 BATSE GRBs were within $45^o$ of its zenith. Milagrito data was analyzed to look for evidence of emission of TeV $\gamma$-rays coincident with the low energy photons detected by BATSE from these GRBs. These GRBs are listed in Table 1. Of these 54 GRBs, there are 12 for which location arcs from the Third Interplanetary Network (IPN3) exist.

## 2 Method of Search

**2.1 Search time duration and search area**  The search time duration used was T90, the time during which BATSE detected from 5% to 95% of the GRB counts. For the sample of 54 GRBs, this time ranged from hundreds of milliseconds to 200 seconds. The radius of the area used for the search was the 95% confidence radius for the BATSE GRB. This confidence radius is given as a function of the statistical error on the GRB position by the BATSE collaboration (Briggs, et al., 1999). The statistical error for the sample ranged from $0.6^o$ to $18.0^o$, corresponding to 95% confidence radii of $6^o$ to $29^o$. Histograms of T90 times and 95% confidence radii are shown in Figures 1a and 1b.

The search area was covered by a grid of non-overlapping rectangular bins of equal areas on the celestial sphere, centered at the GRB (DEC,RA) position given by BATSE (see Figure 2). Optimal bin sizes were used in order to maximize the significance of a signal. These depend on the background count and Milagrito's angular resolution (Schnee, 1996). After this grid was searched, to ensure sensitivity to signals located near bin edges, the grid was shifted by half a bin width in DEC, then half a bin width in RA, then half a bin width in both. A search was done at each of these grid configurations, the end result being a search with overlapping, non-independent bins.

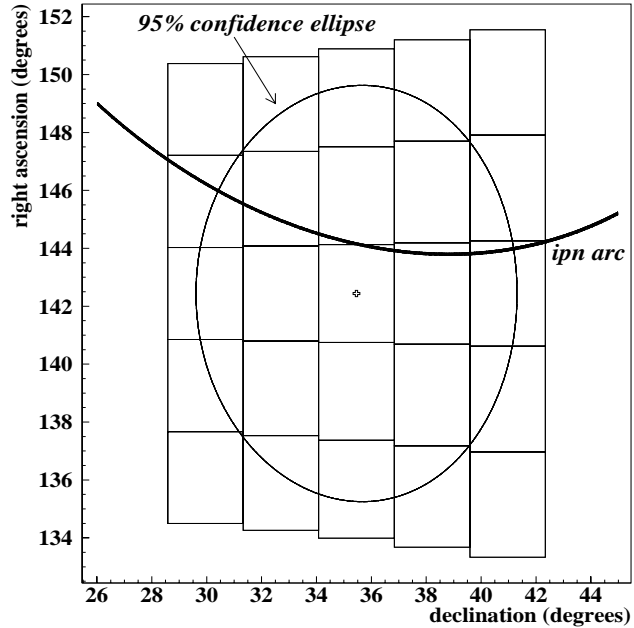

Figure 2: Search area and bin configuration for one GRB, showing the 95% confidence ellipse as well as the IPN3 arc. The center of the ellipse is the nominal GRB position given in the BATSE 4B catalog.

**2.2 Background estimation** Each (DEC,RA) bin of the search area at the middle of the T90 burst time interval (the assumed burst time) was mapped onto a corresponding local bin of the sky, ($\theta,\phi$), and a careful measurement of the background count due to cosmic rays was made for each local ($\theta,\phi$) bin. Events occurring within a six-hour interval centered at the burst time, but excluding events occurring during the T90 burst interval, were put into time bins of duration T90 each. For each time bin, a different piece of the celestial sky was mapped onto the local sky. The events from each time bin falling into each local ($\theta,\phi$) bin were counted. The total number of events, $N_{off}(\theta,\phi)$, falling into a local bin ($\theta,\phi$), divided by the total number of events, $N_{tot}$, coming from the entire sky, is the efficiency of the local bin ($\theta,\phi$) for detecting backround events. The total number of events, $N_{T90}$ occurring during the T90 search interval, and the total number of events, $N_{2min}$, occurring during a two-minute period centered at the middle of the burst search interval were also measured. The background estimate for the local ($\theta,\phi$) bin is then

$$B(\theta,\phi) = \frac{N_{off}(\theta,\phi)}{N_{tot}} N_{search} \qquad (1)$$

where $N_{search} = (N_{2min}/120\ s) \times T90$ for T90 < 120 s, and $N_{search} = N_{T90}$ for T90 > 120 s. The use of $N_{2min}$ is a safeguard against large uncertainties in $N_{T90}$ when T90 is small. The background estimate in the form given by eq. (1) also ensures that rate changes occurring during the six-hour background period do not affect the background estimate for the search period.

**2.3 Determining probabilities** The number of events, $N_{on}$(DEC,RA), falling in each (DEC,RA) bin is measured and the Poisson probability for observing $N_{on}$(DEC,RA) events given an expected background of $B(\theta,\phi)$ at the local bin is calculated. Note that for the short search time intervals used in this analysis, the source was treated as if it did not move relative to the local sky.

# 3 Results

The results of this analysis, including significances of the measurements, will be presented at the conference.

| BATSE trigger number | BATSE trigger date | BATSE fluence, E > 20 keV (ergs/cm$^2$) | Milagrito zenith angle | BATSE trigger number | BATSE trigger date | BATSE fluence, E > 20 keV (ergs/cm$^2$) | Milagrito zenith angle |
|---|---|---|---|---|---|---|---|
| 6100 | 970223 | 7.85E-5 | 24.9$^o$ | 6376 | 970910 | 2.10E-7 | 38.0$^o$ |
| 6128 | 970317 | 4.00E-6 | 43.6$^o$ | 6385 | 970918 | 2.29E-7 | 5.6$^o$ |
| 6129 | 970318 |  | 44.6$^o$ | 6396 | 970925 | 2.83E-6 | 21.9$^o$ |
| 6148 | 970330 | 4.23E-6 | 35.0$^o$ | 6437 | 971015 | 4.56E-7 | 43.4$^o$ |
| 6165 | 970408 | 3.76E-6 | 45.0$^o$ | 6439 | 971016 | 9.58E-8 | 32.1$^o$ |
| 6166 | 970408 | 7.97E-8 | 30.3$^o$ | 6443 | 971021 | 1.07E-6 | 44.9$^o$ |
| 6167 | 970409 | 9.21E-6 | 1.4$^o$ | 6472 | 971110 | 2.67E-4 | 18.1$^o$ |
| 6188 | 970417 | 3.95E-7 | 21.4$^o$ | 6492 | 971122 |  | 43.7$^o$ |
| 6209 | 970426 | 3.37E-6 | 28.6$^o$ | 6523 | 971207 | 2.79E-6 | 43.1$^o$ |
| 6213 | 970429 |  | 25.6$^o$ | 6529 | 971210 | 9.00E-7 | 33.3$^o$ |
| 6219 | 970503 | 1.47E-7 | 38.0$^o$ | 6545 | 971225 | 3.56E-6 | 29.1$^o$ |
| 6229 | 970511 |  | 19.7$^o$ | 6577 | 980124 | 1.96E-6 | 44.7$^o$ |
| 6240 | 970523 | 2.19E-5 | 10.7$^o$ | 6581 | 980125 | 4.88E-5 | 18.2$^o$ |
| 6251 | 970603 | 4.45E-6 | 35.1$^o$ | 6590 | 980207 | 1.16E-5 | 30.1$^o$ |
| 6265 | 970612 | 7.81E-7 | 28.7$^o$ | 6599 | 980213 | 4.22E-6 | 38.1$^o$ |
| 6267 | 970612 | 1.06E-6 | 27.6$^o$ | 6610 | 980222 | 4.36E-6 | 42.8$^o$ |
| 6279 | 970627 | 7.14E-6 | 33.9$^o$ | 6613 | 980223 | 5.91E-7 | 17.1$^o$ |
| 6288 | 970629 | 2.38E-6 | 11.5$^o$ | 6619 | 980301 | 6.16E-6 | 3.7$^o$ |
| 6295 | 970707 | 6.59E-6 | 33.5$^o$ | 6641 | 980315 | 5.94E-7 | 32.4$^o$ |
| 6300 | 970709 | 4.53E-7 | 17.9$^o$ | 6665 | 980329 | 8.26E-5 | 24.1$^o$ |
| 6305 | 970713 | 7.76E-7 | 24.0$^o$ | 6666 | 980329 | 1.48E-6 | 32.9$^o$ |
| 6317 | 970725 | 8.33E-7 | 18.2$^o$ | 6672 | 980401 | 7.81E-6 | 28.2$^o$ |
| 6323 | 970802 | 2.00E-6 | 38.0$^o$ | 6679 | 980404 | 2.62E-6 | 25.7$^o$ |
| 6325 | 970803 |  | 28.3$^o$ | 6694 | 980420 | 2.48E-5 | 34.7$^o$ |
| 6338 | 970817 | 5.46E-7 | 24.2$^o$ | 6702 | 980424 | 1.03E-5 | 44.4$^o$ |
| 6358 | 970903 |  | 36.9$^o$ | 6716 | 980430 | 1.53E-6 | 10.5$^o$ |
| 6366 | 970906 | 1.86E-5 | 44.6$^o$ | 6720 | 980503 | 1.49E-6 | 33.7$^o$ |

Table 1: The 54 GRBs used in this analysis. Blank spaces indicate that no information was found in the BATSE 4B catalog.

*This research was supported in part by the National Science Foundation, the U. S. Dpeartment of Energy Office of High Energy Physics, the U. S. Department of Energy Office of Nuclear Physics, Los Alamos National Laboratory, the University of California, the Institute of Geophysics and Planetary Physics, The Research Corporation, and CalSpace.*